\documentclass[conference]{IEEEtran}
\IEEEoverridecommandlockouts
\usepackage{cite}
\usepackage{url}
\usepackage{microtype}
\usepackage{amsmath,amssymb,amsfonts}
\usepackage{algorithmic}
\usepackage{graphicx}
\usepackage{textcomp}
\usepackage{hyperref}
\usepackage{bbold}
\usepackage{xcolor}
\def\BibTeX{{\rm B\kern-.05em{\sc i\kern-.025em b}\kern-.08em
    T\kern-.1667em\lower.7ex\hbox{E}\kern-.125emX}}
\begin{document}

\title{The Sound of Silence: Efficiency of First Digit Features in Synthetic Audio Detection}

\author{\IEEEauthorblockN{Daniele Mari, Federica Latora, Simone Milani}
\IEEEauthorblockA{\textit{Department of Information Engineering}, \textit{University of Padova}, Padova, Italy \\
\{daniele.mari,federica.latora,simone.milani\}@dei.unipd.it}
}

\maketitle

\begin{abstract}
The recent integration of generative neural strategies and audio
processing techniques have fostered the widespread of synthetic speech synthesis or transformation algorithms. This capability proves to be harmful in many legal and informative processes (news, biometric authentication, audio evidence in courts, etc.). Thus, the development of efficient detection algorithms is both crucial and challenging due to the heterogeneity of forgery techniques.

This work investigates the discriminative role of silenced parts in synthetic speech detection and shows how first digit statistics extracted from  MFCC coefficients can efficiently enable a robust detection. The proposed procedure is computationally-lightweight and effective on many different algorithms since it does not rely on large neural detection architecture and obtains
an accuracy above 90\% in most of the classes of the ASVSpoof dataset.
\end{abstract}

\begin{IEEEkeywords}
synthetic speech detection, first digit statistics, fake audio detection, silenced signal analysis, Random Forest
\end{IEEEkeywords}

\section{Introduction}
Fake human speech recordings have recently proved significantly harmful with respect to misinformation, fake news widespreading, frauds, and ID replacement\cite{ceoscam}. Such results are the outcome of the recent evolution of computing facilities and deepfake technologies, which has allowed the generation of more and more credible synthetic images, videos, and speech audio signals. As a matter of fact, this development has urged the need for accurate fake audio detection strategies that help human listeners in discriminating fraudulent audio samples from bonafide ones.

Several fake audio detection strategies were proposed in literature targeting different types of acoustic features that are present in a real signal and, at the same time, are difficult to synthesize.

Traditional methods rely on estimating fake audio peculiarities from audio transform coefficients like MFCC or LPC \cite{kamble_sailor_patil_li_2020}. More recently, such coefficients have been replaced by learned feature representations generated with CNN or RNN architectures \cite{zhang2017investigation}.
Some other strategies rely on the effects of the physical acquisition environment on the signal (e.g., reverberation, noise, etc.) \cite{6854466,lieto2019hello,capoferri2020speech} or on prosodic and emotional characteristics  \cite{conti2022deepfake}. 
Other solutions rely on statistics and symmetry properties of speech signals \cite{9565457}. Among these, it is worth mentioning the  First Digits (FD) statistics computed on signal transform coefficients\cite{bianchimp3,4512175,4378977}, whose applications have been widely exploited in other multimedia contents \cite{bonettini2021use}. 

Although these solutions aim at detecting the peculiar characteristics of a fake audio signal, more recent works \cite{silencegold} have highlighted how synthetic speech algorithms prove to be effective in spoken parts but fail in generating realistic silence. The work by Muller \emph{et al.} shows that the length of trailing silenced parts\footnote{Silent intervals at the end and at the beginnning of the audio sequence.} in synthetic speech samples from ASVSpoof dataset \cite{yamagishi2021asvspoof} prove to have different statistics with respect to bonafide samples. Indeed, removing such parts dramatically reduces the detection efficiency of most algorithms.

The current paper aims at investigating this eventuality more in depth by analyzing the discriminative potentialities of silenced parts in ASVSpoof dataset. More precisely, we show that FD statistics prove to be effective in discriminating fake audio samples since they allow catching irregularities in silenced parts between the different words of the speech. Tests were run both on the full audio sequence, on the silenced parts, and on the voiced segments (regardless of their lengths). Experimental results show that the statistical characteristics of silence prove to be a discriminative feature since the performance on silent sections matches the detection performance on the full sequence (while this result is not verified for voiced sections). This implies that silence extraction is no longer needed, allowing to avoid parameter tuning and arbitrary set-ups. The final performance has proved to be higher than previous state-of-the-art approaches with a limited computational effort.

It is possible to summarize the novel contributions of the current paper as follows.
\begin{itemize}
    \item We analyzed the role of silent parts in detection showing that most of the classification accuracy derives from the difficulty in synthesizing statistically-realistic silence intervals.
    \item We evaluated the efficiency of MFCC FD statistics in detecting audio fake samples generated by a set of different heterogeneous algorithms. Such features have proved to be extremely useful in highlighting the statistics of silenced parts.
    \item We designed a lightweight classifier whose efficiency can cope with more complex detectors.
\end{itemize}

The code developed to produce the results presented in this work can be found at \href{https://github.com/Dan8991/The-Sound-Of-Silence}{https://github.com/Dan8991/The-Sound-Of-Silence}.

In the following, the paper is organized as follows. Section~\ref{sec:related} overviews some audio forgery detection algorithms that have been proposed in the literature. Section~\ref{sec:algo} describes the proposed approach, Section~\ref{sec: dataset} illustrates the dataset and the experimental setup, while Section~\ref{sec:results} reports the final accuracy on different types of datasets. Final conclusions are drawn in Section~\ref{sec:conclusions}.

\section{Related Works}\label{sec:related}

Generative audio speech approaches can mainly be divided in two branches i.e. text to speech (TTS) and voice conversion (VC) algorithms. The former starts from a textual representation and aims at producing the corresponding waveform, while the latter modifies the signal to change the perceived identity of the speaker in the audio.

Early TTS approaches were based on waveform concatenation \cite{black1995optimising, panda2017waveform} where diphones from large datasets are concatenated seamlessly. More recently researchers have started to design techniques that produce audio features from text representations using an acoustic model (usually a hidden markov model) \cite{reddy2017robust, tokuda2002hmm} that are then processed with a vocoder synthesizer such as STRAIGHT \cite{kawahara2006straight}, WORLD \cite{morise2016world} or VOCAINE \cite{agiomyrgiannakis2015vocaine} to produce the corresponding waveform. To improve upon this, neural networks have also been used to substitute either the acoustic model \cite{wang2016first} or the vocoder \cite{oord2016wavenet, valin2019lpcnet} later leading to the first end to end TTS generation algorithms \cite{wang2017tacotron, ping2017deep}.

On the other hand, VCs pipelines usually extract an intermediate representation of the audio signal (feature extraction step), this is then mapped to a representation that matches the target characteristics (feature mapping step) which is finally used to obtain the final waveform (reconstruction step).

Most feature extraction techniques are usually based on pitch synchronous overlap and add (PSOLA) \cite{arslan1999speaker} that represents the input as the parameters required by a vocoder synthesizer to reproduce it. This is a useful intermediate characterization of the signal because it allows performing reconstruction with a vocoder, which is convenient since these algorithms are well tested and efficient. On the other hand, the mapping function is usually implemented with parallel training methods by using a gaussian mixture model \cite{stylianou1998continuous} or neural networks \cite{ming2016deep, tanaka2019atts2s}. The mapping can also be performed by means of Generative Adversarial Networks (GANs) since the task is similar to image to image translation allowing similar techniques to be adopted \cite{kaneko2018cyclegan, kaneko2019cyclegan}.

Classic audio forgery detection algorithms usually perform classification
by relying on hand crafted features such as Constant-Q Cepstral Coefficients \cite{todisco2017constant}, Log Magnitude Spectrum or phase-
based features like Group Delay \cite{xiao2015spoofing} and Linear Frequency Cepstral Coefficients (LFCC) or MFCCs \cite{sahidullah2015comparison}. More discriminative representations have been recently proposed by exploiting the bicoherence matrix \cite{albadawy2019detecting}, long-short term features computed in an autoregressive manner \cite{borrelli2021synthetic}, environmental cues \cite{capoferri2020speech}, and even emotions \cite{conti2022deepfake}. 

Also in this case neural network based techniques have proven very effective. Some examples are \cite{lieto2019hello}, where the frequency representations of the signals are fed to simple convolutional neural networks (CNNs), and in \cite{zhang2017investigation} where the convolutional filters are just used for feature extraction while a recurrent neural network is exploited for classification. Some approaches have also been directly applied to the raw input signal (i.e. in the time domain) \cite{tak2021end}. In particular, Rawnet2 \cite{tak2021end} has achieved impressive results both for synthetic speech detection and user identification. For this reason, it has been proposed as the baseline for the ASVSpoof 2021 challenge \cite{yamagishi2021asvspoof} i.e. where the dataset considered in this paper for training and testing was proposed.

\section{First digit features for synthetic audio tracks.}\label{sec:algo}

First digit law has proved very effective in the detection of multiple compressed data \cite{7084313,ben12:jpeg,milani_tagliasacchi_tubaro_2014}. More recently, it has also been shown its effectiveness in detecting GAN generated images\cite{bonettini2021use}. Following this trend, it is possible to verify that any synthetic signal generated by a set of FIR filters with limited support fits Benford's law with a different accuracy with respect to a natural signal. 

Audio waveforms $x(t)$ are represented in the frequency domain by computing the MFCC coefficients $m_{w}(f)$, where $f$ is the considered frequency and $w$ is the index of the frame. This representation has already proved very effective in highlighting the more meaningful frequency elements in audio signals and in detecting forged waveforms \cite{sahidullah2015comparison}. 
Since the original samples in the considered dataset sometimes contain long sequences of zeros (which result in zero-valued MFCCs coeffiecients) and since computing FD statistics requires processing non-zero signals, zero values were removed from the input data. This operation does not compromise the final results because this eventuality was verified on both training and test sets, as well as on both natural and synthetic audio.

In order to obtain rich features that can highlight irregularities in the data, MFCC coefficients were quantized with different step values $\Delta$ as 
\begin{equation}
    m_{w, \Delta}(f) = \frac{m_{w}(f)}{\Delta}.
\end{equation} 

At this point, first digits were computed on $m_{w, \Delta}(f) $ as
\begin{equation}
d_{w, \Delta}(f) = \left\lfloor \frac{|m_{w, \Delta}(f)|}{b^{\lfloor log_b|m_{w, \Delta}(f)| \rfloor}} \right\rfloor
\end{equation}
where $b$ is the considered integer representation  base (e.g. 10 for decimal).

For each distinct cepstral coefficient and for each quantization step, we computed the probability mass function
\begin{equation}
    p_{f, \Delta}(d) = \sum_{w=1}^{n_w}\frac{\mathbb{1}_d(d_{w, \Delta}(f))}{n_w}
\end{equation}
where $\mathbb{1}_d(d_{w, \Delta}(f))$ is the indicator function for digit $d$, and $n_w$ is the number of windows in the signal whose value depends on the duration of the audio and on the window overlap.

Several previous studies show that this p.m.f. can be approximated by the generalized Benford's law, i.e.,
\begin{equation}
    \hat{p}_{f, \Delta}(d) = \beta log_b \bigg( 1 + \frac{1}{\gamma + d^\delta} \bigg)
\end{equation}
\noindent and the approximation accuracy highly varies if we are considering bonafide w.r.t. forged data \cite{4512175,4378977}. As a matter of fact, such accuracy was measured using different distance and divergence measures to quantify the proximity of $p_{f, \Delta}(d)$ w.r.t. $\hat{p}_{f, \Delta}(d)$. In the rest of the paper, we will omit indexes $\Delta$ and $f$ for the sake of simplicity although in the creation of the final set of features multiple values of $f$ and $\Delta$ were considered.

A first traditional divergence metric is the Shannon divergence
\begin{equation}
    D^{JS}(p|\hat{p}) = D^{KL}(p|\hat{p}) + D^{KL}(\hat{p}|p).
\end{equation}
\noindent which can be seen as a symmetrized version of the Kullbak-Leibler divergence $D^{KL}(p|\hat{p})$.
Additionally, since such metric proves to be unstable for biased pmfs, so we computed Reny $D_\alpha^R(p|\hat{p}) $ and Tsallis $D_\alpha^T(p|\hat{p}) $  ($\alpha \in [0, 1]$) divergences as well

\begin{equation}
    D_\alpha^R(p|\hat{p}) = \frac{1}{1-\alpha}\big(logS_\alpha(p, \hat{p}) + logS_\alpha(\hat{p}, p)\big)
\end{equation}
\begin{equation}
    D_\alpha^T(p|\hat{p}) = \frac{1}{1-\alpha}\big( 2 - S_\alpha(p, \hat{p}) - S_\alpha(\hat{p}, p)\big)
\end{equation}

where 

\begin{equation}
    S_\alpha(p, q) = \sum_{d=1}^{b-1}\frac{p(d)^\alpha}{q(d)^{\alpha - 1}}
\end{equation}

Additionally, since Reny, Tsallis, and Shannon divergences can be highly correlated for certain values of $\alpha$ (in this work $\alpha = 0.3$ is used), we also added 
the mean square error
\begin{equation}
    D^{MSE}(p, \hat{p}) = \frac{1}{b - 1} \sum_{d=1}^{b-1} (p(d) - \hat{p}(d))^2
\end{equation}
This addition was supported by some preliminary tests where the divergences of original and voice converted audios were compared: it was possible to deduce that Reny Tsallis and Shannon divergences often agree, meaning that the three divergences in the original sample are always smaller than those in the forged sample or vice-versa. This statement does not always hold for MSE. 

In the end,  the total number of features $n_f$ is equal to $n_f = n_d n_c n_b n_q$ where $n_d$ is the number of divergences, $n_c$ is the number of chosen cepstral coefficients, $n_b$ is the number of basis for the first digit extraction and $n_q$ is the number of different $\Delta$ parameters.

\subsection{A FIR-oriented interpretation of FD statistics for synthetic speech}

In the past literature, several works have provided different explanations for the effectiveness of FD statistics in detecting forgeries (on images, audio files, etc.) \cite{4378977}. Most of the proposed works were focusing on the original data statistics on which FDs were computed. Indeed, Benford's law and its generalized version can be verified for any set of data $m$ such that their probability mass function (pmf) has an exponentially-decreasing behavior (this has been largely verified on images, where coefficients can be modeled with a Laplacian or a generalized Gaussian distribution)\cite{4512175}. Whenever the image or the set of data are altered, the property is not verified anymore since the modification redistributes data among the bins of the quantizer. Indeed, the final pmf presents some oscillating probability values that deviate from the ideal distribution. 

\begin{figure}[t]
\begin{tabular}{c}
\begin{minipage}[c]{0.7\columnwidth}
\includegraphics[width=\columnwidth]{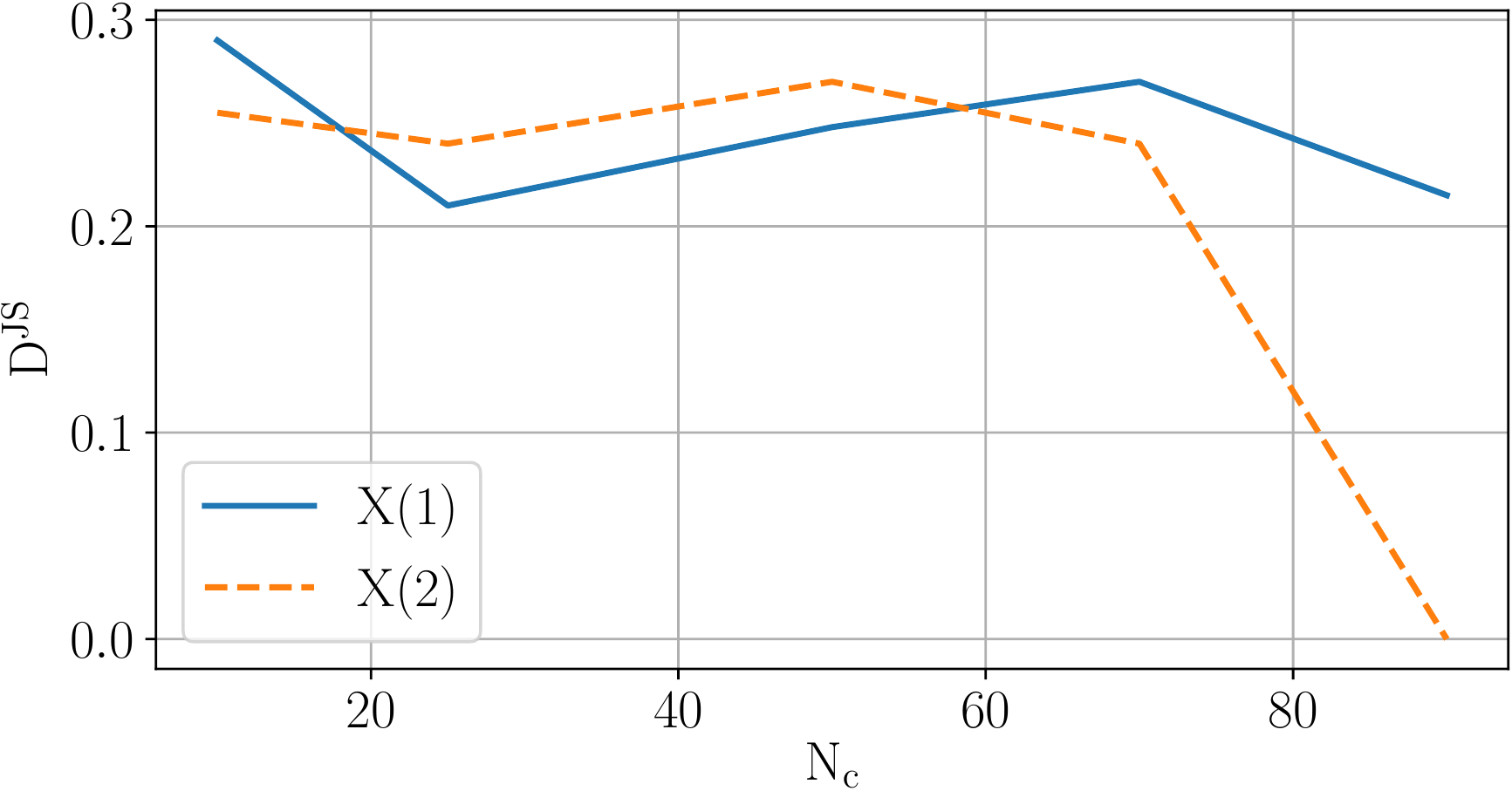}
\end{minipage} \\ \centerline{(a)} \\
\begin{minipage}[c]{0.7\columnwidth}
\includegraphics[width=\columnwidth]{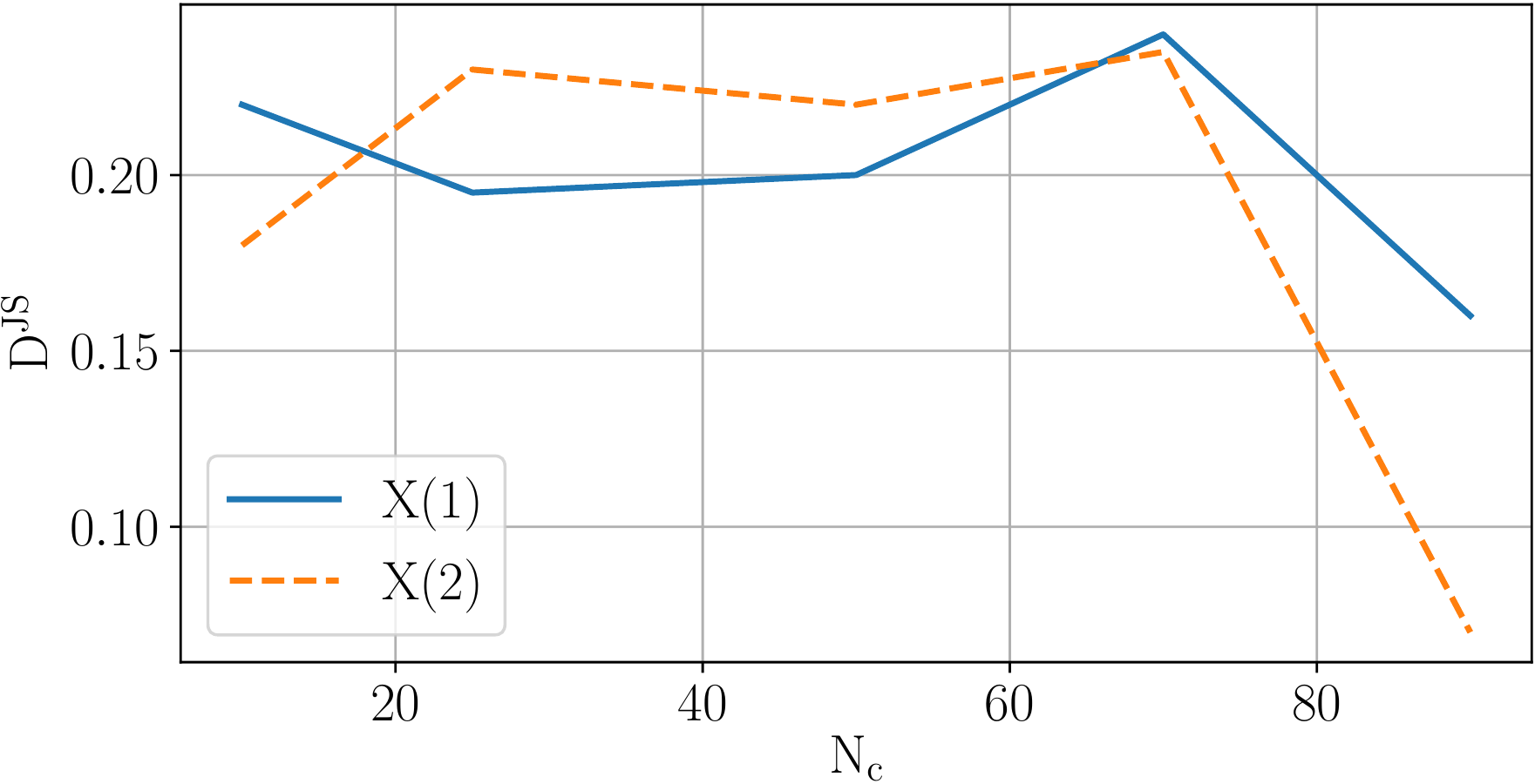}
\end{minipage} \\
\centerline{(b)}
\end{tabular}
\caption{Jensen-Shannon divergence values between computed and fitted \textit{fd} statistics. Data were computed with (a) $\Delta=0.008$ and (b) $\Delta=0.01$.}\label{fig:div}
\vspace{-1.5em}
\end{figure}

 Anyway, such oscillations can be also related to the ripples in the frequency response of small FIR filters (like those used in GAN-based or VOCODER-based speech synthesizers) that propagates to the statistics of MFCC coefficients. After these multiplications,  the values $|m_{w,\Delta}(f)|$ belonging to the same quantization bin are re-distributed unevenly among the other bins. Instead, whenever the gain is perfectly flat, all the MFCC coefficients are rescaled with the same factor, and as a matter of fact, the whole statistics is simply stretched.
From these considerations, it is possible to conclude that the flatness of filter gain is deeply connected to the verification of Benford's law. Fig.~\ref{fig:div}a,~\ref{fig:div}b report the values of  the Jensen-Shannon divergence $D^{JS}\left(\hat{p}_{f,\Delta}(d)|p_{f,\Delta}(d) \right)$ between $\hat{p}_{f,\Delta}(d)$ and $p_{f,\Delta}(d)$ obtained on a $1$-D Gaussian i.i.d. signal $x(t)$ filtered by an FIR filter $h(t)$. The filter $h(t)$ is a standard low-pass FIR filter with normalize cut-off frequency $0.2$, stop-band frequency $0.7$, and $N_c$ coefficients. Filter coefficients were computed using the Parse-McLellan algorithm.

Divergences have been computed from the statistics of quantized MFCC coefficients computed at frequencies $2$ and $3$. It is possible to notice that $D^{JS}\left(\hat{p}(d)|p(d) \right)$ decreases as $N_c$ increases; however, such behavior can be more or less enhanced depending on coefficient frequency and the quantization parameter $\Delta$. As a matter of fact, it is necessary to use different frequency values and quantization set-ups. 

\section{Dataset preparation and experimental setup}
\label{sec: dataset}
The dataset used to validate the proposed approach is ASVSpoof \cite{yamagishi2021asvspoof} since it provides a great variety of synthetic speech samples generated by heterogeneous algorithms (see Tables~\ref{tab:on set},\ref{tab:off set}). In fact, ASVSpoof covers both text to speech (TTS) and voice conversion (VC) scenarios including samples generated by algorithms based on waveform concatenation (WC), transfer function (TF), non parallel voice conversion systems (NP), and neural networks (NN).

Data can be divided in three main parts: training, development, and evaluation datasets. The former was used in training and selecting the final classification model. Samples were randomly decimated to ensure that every generated class has the same number of audio traces and that the total amounts of bonafide and synthetic data are balanced (the adopted classifier is a random forest which has lower overfitting problems so the data used for training is plenty).

Development and evaluation datasets were used for closed-set and open-set testing, respectively. More precisely, the former includes newly generated samples (not seen before) generated by the same set of  algorithms of the training set, while the latter includes samples generated by different strategies not included in the training set.

The work by Muller et. al. \cite{silencegold} shows that a bias can be found in the distribution of the lengths of leading and trailing silences in bonafide and synthetic speeches. Authors argue that most detectors are just probably discriminating between forged and bonafide samples by using this information. In order to bypass this problem, silent parts were removed from the signal, as suggested in \cite{silencegold} but this led to a big loss in performance. 

Therefore, we have decided to analyze the effectiveness of FD features on the silent (without considering leading and ending silences) and voiced parts of the signals, independently. This allowed us to understand which speech elements proved to be the most discriminative and whether the proposed approach was reliably effective. For this purpose, we selected signal windows of $101$ samples with energy $E(s, t)$ higher than $-40$ dB (assuming energy is normalized). 

From this filtering, only a few samples (less than 1\%) were then removed since the number of silent values was not enough to obtain meaningful statistics.  Arguably, this is not an issue since as shown in \cite{silencegold}, the very low amount of silence in the audio track allows an easy detection of synthetic audio samples. Moreover, computing FD statistics on a limited amount of signal windows would lead to highly irregular statistics: this implies strong divergences/distances with respect to Benford's law (and therefore, a correct classification). 

Starting from the original samples, three datasets have been generated, one called \textit{Full} containing the whole waveform, one referred as \textit{Silence} made with the silent parts of the signals, and one called \textit{Voiced} with the remaining samples.

On these samples, cepstral analysis was carried on in order to generate a feature array for each sample. In this process, the following parameter values were selected after an extensive set of optimizations.
\begin{itemize}
    \item In the computation of MFCCs, a filter bank of 26 filters was adopted: only coefficients from the second to the fourteenth frequency were considered. Computation was carried out on window sizes of 1024 samples with an overlap of 512 in the case of \textit{Full} and \textit{Voiced}. Overlap was set to 128 in the case of \textit{Silence} to have a sufficient number of signal windows (and therefore stable FD statistics).
    \item The base for the first digit was chosen as $b \in \{10, 20\}$ since higher values would imply only a few samples (or no samples at all) for many FD values.
    \item The quantization factor $\Delta$ varied in the set $\{1, 2, 3, 4\}$.
\end{itemize}

At the end of the generation process, feature arrays were made of $n_f = 420$ features.

Given the number and the statistical independence of features (as well as the need for a low complexity classifier), we avoided the adoption of complex neural network architectures. For this reason, a simple random forest classifier was selected as it proved well suited for tabular data processing and highly robust w.r.t. overfitting problems and unbalancing.

The best configuration was selected by running a grid search over the number of trees in the random forest ($n_{trees} \in \{10, 100, 500, 1000\}$) and the criterion for the split quality ($criterion \in \{gini, entropy\}$). 

\begin{table}[]
    \centering
    \begin{tabular}{|c|c|c|c|c|c|c|}
        \hline
        \multicolumn{1}{|c|}{\textbf{Dataset}} &\multicolumn{6}{c|}{\textbf{Development}}\\ 
        \hline
        \textbf{Algorithm} &\textbf{A01} & \textbf{A02} & \textbf{A03} & \textbf{A04} & \textbf{A05} & \textbf{A06} \\
        \textbf{Type} & TTS & TTS & TTS & TTS & VC & VC\\
        \textbf{Approach} & NN & NN & NN & WC & NN & TF\\
        \hline
        \textbf{Silence $\Delta$=1} & 0.944 & 0.962 & 0.961 & 0.819 & 0.949 & 0.471\\
        \textbf{Silence $\Delta$=1-2} & \textbf{0.953} & 0.972 & 0.970 & 0.829 & 0.961 & 0.472\\
        \textbf{Silence $\Delta$=1-3} & 0.951 & 0.972 & 0.972 & 0.836 & \textbf{0.964} & 0.466\\
        \textbf{Silence $\Delta$=1-4} & 0.952 & 0.973 & 0.972 & 0.838 & 0.963 & 0.456\\
        \textbf{Silence b=10} & 0.945 & 0.959 & 0.961 & 0.830 & 0.924 & 0.468\\
        \textbf{Silence b=20} & 0.866 & 0.973 & 0.881 & 0.796 & 0.957 & 0.434\\
        \textbf{Full $\Delta$=1-4} & 0.951 & \textbf{0.982} & \textbf{0.949} & \textbf{0.871} & 0.956 & 0.424\\
        \textbf{Voiced $\Delta$=1-3} & 0.755 & 0.708 & 0.713 & 0.548 & 0.574 & \textbf{0.532}\\
        \hline
    \end{tabular}
    \vspace*{0.5em}
    \caption{On-set results and ablation studies for the proposed algorithm}
    \label{tab:on set}
    \vspace{-3em}
\end{table}

\begin{table*}[]
    \centering
    \begin{tabular}{|c|c|c|c|c|c|c|c|c|c|c|c|c|c|}
        \hline
        \multicolumn{1}{|c|}{\textbf{Dataset}} &\multicolumn{13}{c|}{\textbf{Evaluation}}\\
        \hline
        \textbf{Algorithm} & \textbf{A07} & \textbf{A08} & \textbf{A09} & \textbf{A10} & \textbf{A11} & \textbf{A12} & \textbf{A13} & \textbf{A14} & \textbf{A15} & \textbf{A16} & \textbf{A17} & \textbf{A18} & \textbf{A19}\\
        \textbf{Type} & TTS & TTS & TTS & TTS & TTS & TTS & TTS+VC & TTS+VC & TTS+VC & TTS & VC & VC & VC \\
        \textbf{Approach} & NN & NN & NN & NN & NN & NN & NN & NN & NN & WC & NN & NP & TF \\
        \hline
        \textbf{Silence $\Delta$=1} & 0.946 & 0.948 & 0.955 & 0.947 & 0.947 & 0.952 & 0.953 & 0.931 & 0.876 & 0.860 & 0.597 & 0.615 & 0.592 \\
        \textbf{Silence $\Delta$=1-2}  & 0.953 & 0.953 & 0.965 & 0.954 & 0.956 & 0.959 & 0.960 & 0.939 & \textbf{0.888} & 0.861 & 0.598 & 0.626 & 0.597 \\
        \textbf{Silence $\Delta$=1-3} & \textbf{0.955} & \textbf{0.957} & \textbf{0.968} & \textbf{0.956} & \textbf{0.957} & \textbf{0.962} & \textbf{0.962} & 0.941 & \textbf{0.888} & 0.864 & 0.600 & 0.629 & \textbf{0.599} \\
        \textbf{Silence $\Delta$=1-4} & 0.951 & 0.955 & 0.965 & 0.952 & 0.956 & 0.960 & 0.959 & \textbf{0.942} & 0.887 & 0.864 & \textbf{0.601} & 0.625 & 0.598\\
        \textbf{Silence b=10} & 0.925 & 0.933 & 0.944 & 0.927 & 0.929 & 0.936 & 0.928 & 0.912 & 0.860 & 0.846 & 0.598 & \textbf{0.642} & \textbf{0.599}\\
        \textbf{Silence b=20} & 0.919 & 0.897 & 0.929 & 0.924 & 0.924 & 0.903 & 0.945 & 0.889 & 0.842 & 0.820 & 0.590 & 0.579 & 0.593\\
        \textbf{Full $\Delta$=1-4} & 0.941 & 0.942 & 0.952 & 0.939 & 0.940 & 0.915 & 0.951 & 0.896 & 0.853 & \textbf{0.866} & 0.597 & 0.581 & 0.596\\
        \textbf{Voiced $\Delta$=1-3} & 0.656 & 0.796 & 0.798 & 0.629 & 0.648 & 0.628 & 0.687 & 0.720 & 0.709 & 0.640 & 0.526 & 0.533 & 0.580\\
        \hline
    \end{tabular}
    \vspace*{0.5em}
    \caption{Off-set results and ablation studies for the proposed algorithm}
    \label{tab:off set}
    \vspace{-2.1em}
\end{table*}

\section{Results}
\label{sec:results}

\begin{table}[]
    \centering
    \begin{tabular}{|c|c|c|}
        \hline
        \textbf{Algorithm} & \textbf{Development} & \textbf{Evaluation} \\
        \hline
        \textit{Silence} & 0.869 & 0.819 \\
        \textit{Full} & 0.871 & \textbf{0.820} \\
        STLT + Bicoherence 128 & \textbf{0.942} & 0.735\\
        STLT + Bicoherence 512 & 0.907 & 0.741\\
        \hline
    \end{tabular}
    \vspace*{0.5em}
    \caption{Classification accuracy of the proposed algorithm and of some state of the art approaches}
    \label{tab:total accuracies}
    \vspace{-3em}
\end{table}

Given the three testing scenarios (\textit{Full}, \textit{Voiced}, \textit{Silence}), various ablation studies were carried out to verify the efficiency of the classification in the different set-ups and identify the most discriminative elements in the FD divergences. In order to guarantee a fair comparison between the various configurations, we run an independent grid search for each features configuration.

Table~\ref{tab:on set} reports the one-vs-one on-set results obtained by performing binary classification between bonafide and samples generated with a single algorithm. Table~\ref{tab:off set} reports the one-vs-one off-set accuracy.

Full ablation studies are reported only for \textit{Silence} for the sake of conciseness, while for \textit{Full} and \textit{Voiced} only the best results are reported. Considering the impact of base selection, keeping the features generated by both $b=10$ and $b=20$ (with all the selected values of $\Delta$) turned out to be more effective than choosing only one base value. Indeed, we verified that the statistics generated for $b=10$ are not very correlated with those obtained for $b=20$, and therefore, merging them provides additional information to the system.

With respect to the quantization parameter $\Delta$, in Table~\ref{tab:on set} and~\ref{tab:off set}  the features related to different $\Delta$s were incrementally concatenated one at a time in order to measure how much they affected the final performance. It is possible to see that having only one  $\Delta$ value is usually not enough to maximize performance. Experiments show that in general $3$ or $4$ different quantization values allow to achieve the best performance.

In both on-set and off-set accuracies, it is possible to see that performing classification over the features computed on \textit{Silent} leads to performance that is comparable to or even better w.r.t. the one obtained for \textit{Full}. In particular, when considering off-set tests, silences provide a higher detection accuracy for almost all the algorithms.

On the other hand, removing silences from the signals leads to very poor performance (see results on \textit{Voiced} sections). This might suggest that algorithms reconstruct realistic voices more easily (low-pass regular signal), while the noise present in silent sequences can not be easily modeled. The slightly lower performance achieved on \textit{Full} w.r.t. \textit{Silence} might be explained by the presence of spoken parts that might be skewing the FD statistic towards the ideal FD distribution. 

It is possible to see that this approach has a lower performance when referred to algorithms A06, A17, A18, A19.  It is worth noticing that all these approaches perform a voice conversion task starting from real audio samples as input and converting them into voiced samples for a desired speaker. On the contrary algorithm A05 is also a VC algorithm but the task is carried out by a neural network that processes the full sequence (silence included) leading to FD statistics that are detected by our approach. Additionally, voice converted samples generated starting from TTS outputs (see results referred to algorithms A13, A14, and A15) are also easily classified with high accuracy.
A possible explanation for this evidence is that VC algorithms do not change significantly silenced sections as they are not relevant in characterizing speaker ID and present a completely different statistics w.r.t. voiced parts. In these cases, the statistics of the original silenced intervals are not altered leading to a higher misclassification probability. Note that this outcome is not verified whenever VC is applied after TTS since in that case also the generated nature of silence leads to non-conventional FD statistics (thus leading to higher divergences/distances).

On top of that, it is worth spending a few comments on A16 and A04 approaches. These are based on waveform concatenation, i.e., signals are obtained by concatenating real samples from big databases of diphones (realistic ones). Such composite nature makes the synthetic speech FD statistics closer to that of bonafide samples (and further from the generated audio) leading to a higher misclassification probability.

In the end, we compared our approach with a similar state-of-the-art algorithm (i.e. with separate feature extraction and classification steps). The work by Borrelli et. al. \cite{borrelli2021synthetic} exploits short-term and long-term (STLT) cues and the bicoherence matrix to extract a discriminative representation between forged and bonafide samples. In Table~\ref{tab:total accuracies} the results of the two best configurations proposed in the aforementioned work, i.e. using both STLT and bicoherence features computed with window sizes 128 and 512, are compared with the results obtained with the best configurations in the \textit{Silence} and \textit{Full} datasets. It is possible to see that while the proposed approach is less accurate in the detection of forged data in the development dataset, it achieves better performance in off-set evaluation proving more robust when presented with unseen algorithms.

\section{Conclusions}\label{sec:conclusions}
In this paper, we analyzed the impact of voiced and silenced parts in synthetic speech detection. Following some preliminary studies on trailing silences, we showed that silenced parts within the speech contain most of the discriminative information. From these results, we proposed a method for forged audio detection based on first digit statistics that achieves good detection performance against a variety of algorithms and that has very low computational complexity. Empirical results showed that most audio forging algorithms are able to produce statistically meaningful voice signals but (especially neural networks) often fail at creating realistic silences. Future works should try to tackle the problem of detection in a voice conversion scenario (possibly by integrating this with other well working state of the art approaches) since the transformation of a naturally acquired signal could retain most of the statistics for the silenced parts thus leading to a higher misclassification probability.

\bibliographystyle{IEEEbib}
\bibliography{biblio}
\end{document}